\documentclass[runningheads]{llncs}
\usepackage{graphicx}
\usepackage{amsmath}
\usepackage{multirow}
\usepackage{wasysym}
\usepackage{booktabs}

\begin{document}
%
\title{Flow-based Deformation Guidance for Unpaired Multi-Contrast MRI Image-to-Image Translation}
%
%

\newcommand\correspondingauthor{\thanks{Corresponding author}}
\author{Toan Duc Bui \correspondingauthor \inst{1} \and
Manh Nguyen \inst{1,2}\and
Ngan Le \inst{3} \and
Khoa Luu \inst{3}}
\index{Bui, Toan Duc} 
\index{Nguyen, Manh}
\index{Le, Ngan}
\index{Luu, Khoa}
\institute{VinAI Research, Vietnam \and 
FPT University \and
Department of Computer Science at University of Arkansas in Fayetteville \\ \email{\{v.toanbd1\}@vinai.io}}

\maketitle              
\begin{abstract}
 
Image synthesis from corrupted contrasts increases the diversity of diagnostic information available for many neurological diseases. Recently the image-to-image translation has experienced significant levels of interest within medical research, beginning with the successful use of the Generative Adversarial Network (GAN) to the introduction of cyclic constraint extended to multiple domains. However, in current approaches, there is no guarantee that the mapping between the two image domains would be unique or one-to-one. In this paper, we introduce a novel approach to \textbf{unpaired image-to-image translation} based on the \textbf{invertible architecture}. The invertible property of the flow-based architecture assures a cycle-consistency of image-to-image translation without additional loss functions. \textbf{We utilize the temporal information between consecutive slices to provide more constraints to the optimization for transforming one domain to another in unpaired volumetric medical images}. To capture temporal structures in the medical images, we explore the displacement between the consecutive slices using a deformation field. In our approach, the deformation field is used as a guidance to keep the translated slides realistic and consistent across the translation.  The experimental results have shown that the synthesized images using our proposed approach are able to archive a competitive performance in terms of mean squared error, peak signal-to-noise ratio, and structural similarity index when compared with the existing deep learning-based methods on three standard datasets, i.e. HCP, MRBrainS13 and Brats2019. 

\end{abstract}

\section{Introduction}
In medical imaging, the task of obtaining diagnostic images from multiple modalities is necessary for accurate and comprehensive prediction of disease diagnosis. For example, T1-weighted (T1) brain images provide clear differentiate images of gray and white matter tissues, whereas T2-weighted (T2) images differentiate fluid from cortical tissue. By leveraging the information provided by both of these image modalities, we can gain a more in-depth and completed picture of the diagnosis. However, obtaining separately both images is often costly, time-consuming, and maybe corrupted by noise and artifacts. Therefore, cross-modalities synthesis is a promising application to improve the clinical feasibility and utility of multi-contrast MRI. Image-to-image translation has recently gained attention in the medical imaging community, where the task is to estimate the corresponding image in the target domain from a given source domain image of the same subject. Generally, the image-to-image translation methods can be divided into two categories including: Generative Adversarial Networks (GANs) and Flow-based Generative Networks and summarized as follows:

\textbf{Generative Adversarial Networks 
GANs} are a class of latent variable generative models that clearly identify the generator as \textit{deterministic mapping}. The deterministic mapping represents an image as a point in the latent space without regarding its feature ambiguity. Several different GAN-based models have been used to explore image-to-image translation in a literature study \cite{bansal2018recycle,chen2019one,welander2018generative,zhu2017unpaired}. For example, Zhu et al. \cite{zhu2017unpaired} proposed a cycleGAN method for mapping between unpaired domains by using cycle-consistency dependence to constrain the optimal solutions provided by the generative network. Balakrishnan et al. \cite{bansal2018recycle} proposed a RecycleGAN to explore the temporal information by learning a prediction of the next frame for video generation. Chen et al. \cite{chen2019one} proposed a 3D cycleGAN network to learn the mapping between CT and MRI. The drawback of 3D cycleGAN is it is memory consumption and loses the global information due to working on small patch sizes.

\textbf{Flow-based Generative Networks} are a class of latent variable generative models that clearly identify the generator as an \textit{invertible mapping}. The invertible mapping provides a distributional estimation of features in the latent space. Recently, many efforts making use of flow-based generative networks have been proposed to transfer between two unpaired data \cite{dinh2016density,grover2019alignflow,kingma2018glow,van2019reversible,sun2019dual}. For example, Grover et al. \cite{grover2019alignflow} introduced a flow to flow (alignflow) network for unpaired image-to-image translation. Sun et al. \cite{sun2019dual} introduced a conditional dual flow-based invertible network to transfer between positron emission tomography (PET) imaging and magnetic resonance imaging (MRI) images. By using invertible properties, the flow-based methods can ensure exact cycle consistency in translation from a source domain to the target and returning to the source domain without any further loss functions.

\textbf{Limitations of Existing Methods and Our Contributions} 
The primary drawback of the cycleGAN model is that it can not perform one-to-one mapping for accurate and unique unpaired image translation, generates biased image translations of the inverse mapping \cite{shen2020one}.  Different from the GANs-based method, the flow-based method guarantees precise cycle consistency in mapping data points from a source domain to the target and returning to the source domain. However, the flow-based methods do not take into account the temporal information between consecutive slices. To address this problem, we propose a new method by inheriting the merits of the flow-based method and exploiting temporal information between consecutive slices. Our approach provides more constraints to the optimization for transforming one domain to another domain. To capture temporal information, we employ a deformation field between consecutive slices by training a convolutional neural network. In our proposed approach, the deformation field plays a role of guidance to keep slices realistic and consistent across translation. 

\section{Related work}

\subsection{Cycle-Consistent Adversarial Networks (cycleGAN)}
Let $\{x_i\}_{i=1}^N$ and $\{y_i\}_{i=1}^M$ be unpaired data samples for two domains, i.e. the source domain $X$ and the target domain $Y$, respectively. Denote $D$ and $G$ as a discriminator network and a generator network. The cycleGAN model \cite{zhu2017unpaired} solves unpaired image-to-image translation between these two domains by estimating two independent mapping functions $G_{X \to Y}: X\to Y$  and $G_{Y \to X}: Y\to X$. The two mapping functions $G_{X \to Y}$ and $G_{Y \to X}$ performed by neural networks are trained to fool the discriminator $D_X$ and $D_Y$ respectively. The discriminator $D_X$, and $D_Y$ encourage the transferred images and the real images to be similar. Hence, the cycleGAN loss is defined as:


\begin{equation}
\begin{split}
 &\mathcal{L}_{cycleGAN}(G_{X \to Y},G_{Y \to X},D_X,D_Y) = \mathcal{L}_{GAN}(G_{X \to Y},D_Y) +\mathcal{L}_{GAN}(G_{Y \to X},D_X) \\
 &\quad\quad\quad\quad\quad\quad\quad\quad+\lambda\mathcal{L}_{cycle}(G_{X \to Y},G_{Y \to X}) +\beta\mathcal{L}_{identity}(G_{X \to Y},G_{Y \to X}) 
\end{split}
\label{eq:loss_cyclegan}
\raisetag{13pt}
\end{equation}
where $\mathcal{L}_{GAN}$ is a GAN loss for the $D$ network \cite{zhu2017unpaired}. $\mathcal{L}_{cycle}$ is a cycle consistency loss that guarantees the transferred image from a time-point is able to bring back to the original image after appearance translation by the generator network $G$. For example, the cycle consistency loss of the data translated from $X \to Y$ via $G_{X}$ and mapped back to the original domain $X$ via $G_{Y}$ is defined as:
\begin{equation}
    \mathcal{L}_{cycle}(G_{X \to Y},G_{Y \to X})=\left\lVert G_{Y \to X}(G_{X \to Y} (x)) - x \right\rVert_{1} 
    \label{eq:loss_cycle_consistency}
\end{equation}

The identity loss $\mathcal{L}_{identity}$ is to regularize the generator to be near an identity mapping when real samples of the target domain are given as the input to the generator. The $\lambda$ and $\beta$ control the contribution of the two objective functions.

\subsection{Flow-based Generative Models}
Flow-based Generative Models are a class of latent variable generative models that clearly identify the generator as an invertible mapping $h: Z \to X$ between a set of latent variables $Z$ and a set of observed variables $X$. Let $p_X$ and $p_Z$ indicate the marginal densities given by the model over $X$ and $Z$, respectively. Using the change-of-variables formula, these marginal densities are defined as
\begin{equation}
p_{X}(x) = p_{Z}(z) \bigg\lvert \det \frac{\partial h^{-1}}{\partial X}\bigg\rvert_{X=x}
\label{eq:prob_z}
\end{equation}
where $z=h^{-1}(x)$ because of the invertibility constraints. In particular, we use a multivariate Gaussian distribution $p_{Z}(z) = \mathcal{N} (\mu,\,0,  \,\textbf{I})$. Unlike adversarial training, flow models trained with maximum likelihood estimation (MLE) explicitly require a prior $p_{Z}(z)$ with a tractable density to evaluate model likelihoods using the change-of-variables formula in the equation (\ref{eq:prob_z}). 

Based on flow-based method \cite{dinh2016density}, Grover et al. \cite{grover2019alignflow} proposed an alignflow method for unpaired image-to-image translation. In the method, the  mapping between two domains $X \to Y$ can be represented through a shared feature space of latent variables $Z$ by the composition of two invertible mapping \cite{grover2019alignflow}:
\begin{equation}
G_{X \to Y} = G_{Z \to Y} \circ G_{X \to Z},  \quad\quad\quad
G_{Y \to X} = G_{Z \to X} \circ G_{Y \to Z}
\label{eq:gen}
\end{equation}
where $G_{X \to Z}= G_{Z \to X}^{-1}$ and $G_{Y \to Z}= G_{Z \to Y}^{-1}$. Due to the fact that composition of invertible mappings is invertible, both
$G_{X \to Y}$ and $G_{Y \to X}$ are invertible \cite{grover2019alignflow}. On the other hand, we can obtain $G_{X \to Y}^{-1} = G_{Y \to X}$. Thus the equation (\ref{eq:loss_cycle_consistency}) can rewrite as 

\begin{equation}
\begin{aligned}
    \mathcal{L}_{cycle}(G_{X \to Y},G_{Y \to X})&=\left\lVert G_{Y \to X}(G_{X \to Y} (x)) - x \right\rVert_{1} \\
    &=\left\lVert G_{X \to Y}^{-1}(G_{X \to Y} (x)) - x \right\rVert_{1}
    = 0
    \label{eq:loss_cycle_consistency_idt}
\end{aligned}
\end{equation}
where  $G_{X \to Y}^{-1}G_{X \to Y}$ results in an identical matrix.

Equation. \ref{eq:loss_cycle_consistency_idt} implies that the  flow-based  methods  can  guarantee  precise  cycle  consistency  in mapping from a source domain to the target and returning to the source domain without additional loss functions. Hence, the alignflow objective loss is defined as:
\begin{equation}
\begin{split}
\mathcal{L}_{flow}(G_{X \to Y},G_{Y \to X},D_X,D_Y)&= \mathcal{L}_{GAN}(G_{X \to Y},D_Y) 
+\mathcal{L}_{GAN}(G_{Y \to X},D_X) \\
& - \lambda_{X}\mathcal{L}_{MLE}(G_{Z \to X})- \lambda_{Y}\mathcal{L}_{MLE}(G_{Z \to Y}) 
\end{split}
\label{eq:loss_flow}
\raisetag{13pt}
\end{equation}
where $\lambda_{Y}, \lambda_{Y} \ge 0$ are hyperparameters that control the importance of the MLE terms for domains $X$ and $Y$ respectively. 
\begin{figure}[b]
\centering
\includegraphics[width=0.65\textwidth]{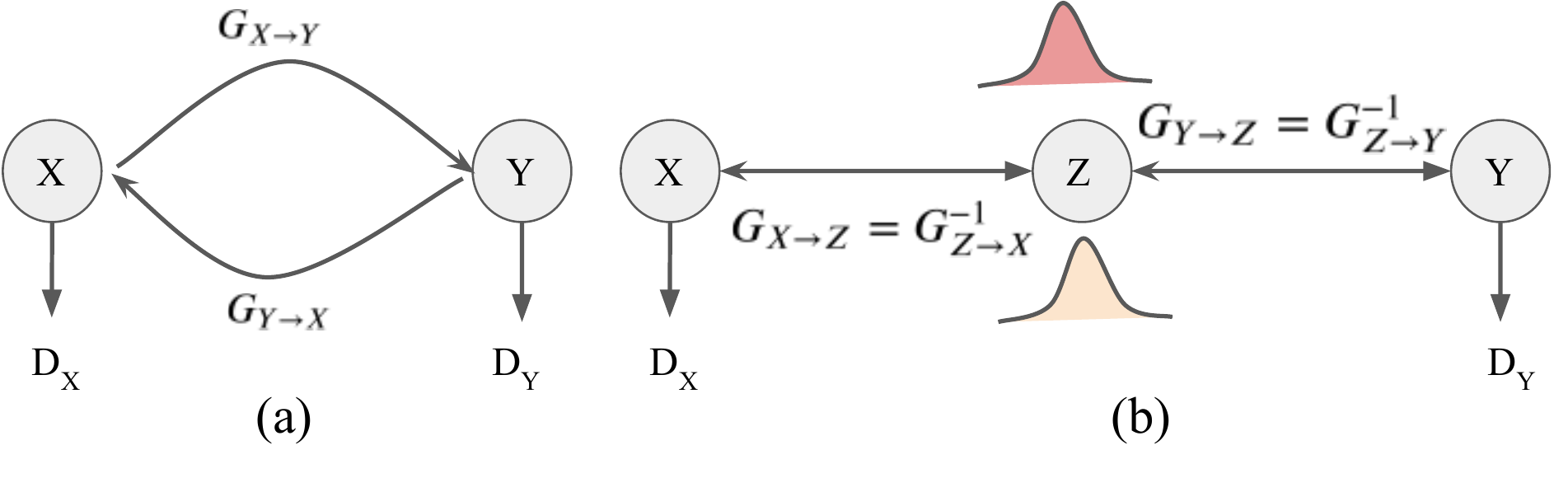}
\caption{A comparison between (a) cycleGAN  and (b) alignflow generative model. Double-headed arrows denotes an invertible mapping}
\label{fig_cyclegan_and_flow}
\vspace{-0.2em}
\end{figure}

Fig. \ref{fig_cyclegan_and_flow} illustrates the difference between cycleGAN and alignflow methods. Unlikes cycleGAN, the alignflow method is the full invertible architecture that guarantees the cycle-consistency translations between two unpaired domains without an additional $\mathcal{L}_{cycle}$ function.

\section{Proposed method}
Our motivation is to learn a mapping between unpaired images from different domains by leveraging the temporal information between consecutive slices. We use the temporal information to constrain the mapping between two domains which should be consistent. Our method is an extension of alignflow \cite{grover2019alignflow} method with making use of temporal information between consecutive slides.

\subsection{Deformation Guided Temporal Constraints}
\begin{figure}[t]
\centering
\includegraphics[width=0.55\textwidth]{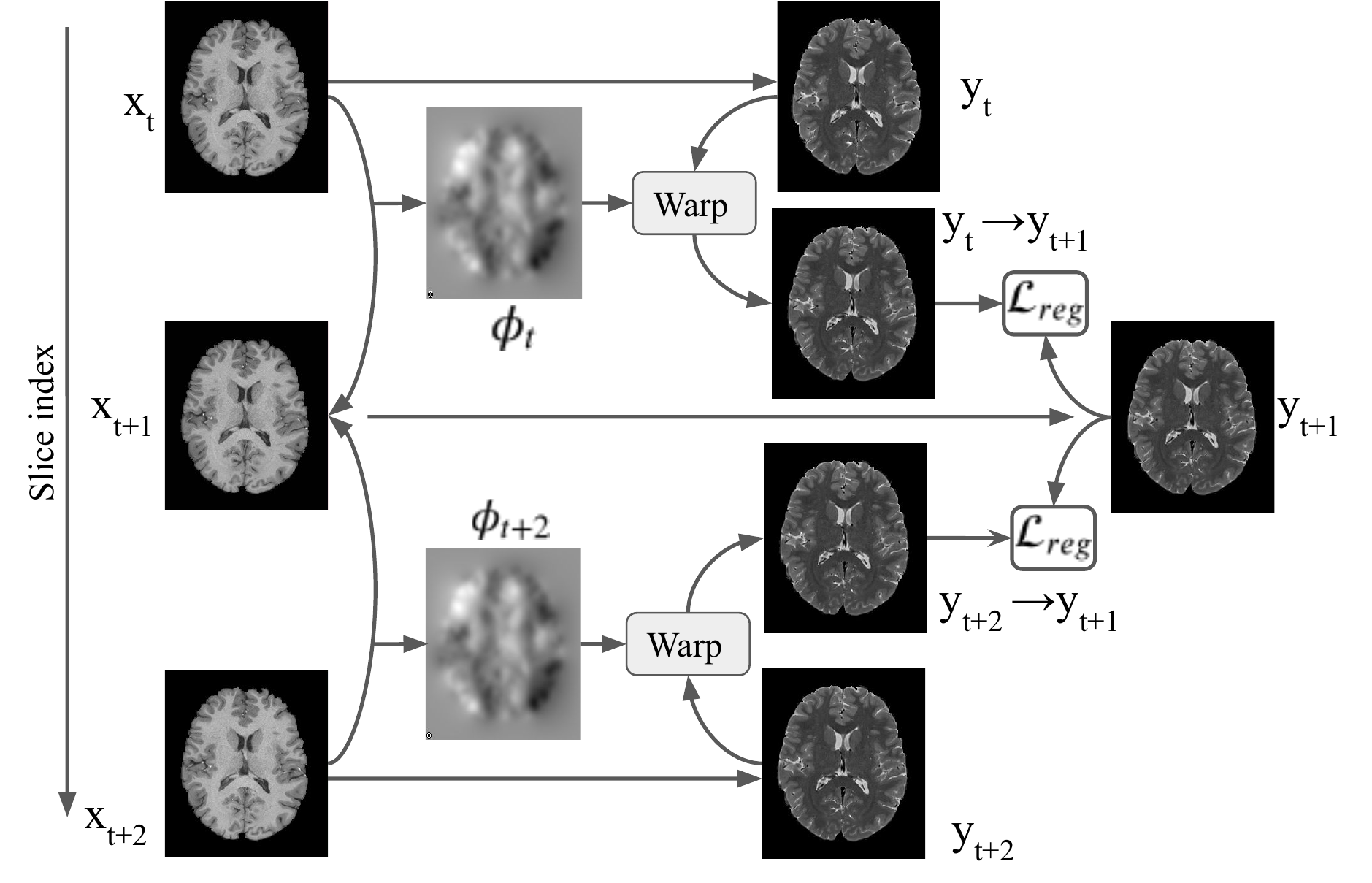}
\caption{Deformation Guided Temporal Constraints for domain $Y$}
\label{fig_deformation}
\end{figure}
To obtain the displacement between consecutive slices, we use an unsupervised registration network \cite{balakrishnan2018unsupervised} to learn a deformation field $\phi$ of a slice $x_t$ and its consecutive slices $x_k$. The deformation field $\phi$ can be obtained using a convolutional neural network (CNN) \cite{balakrishnan2018unsupervised} by minimizing the loss function
\begin{equation}
\mathcal{L}(\phi) =  \lVert x_t - \big(x_k \ocircle \phi(x_t, x_k)\big)\rVert_2  + \lVert \nabla \phi \rVert_2
\label{eq:loss_phi}
\end{equation}
where $\ocircle$ denotes the spatial transformation operation. The first term ensures that the distance between the next slice $x_t$ and the warped current slice $x_k \ocircle \phi(.)$ to be close. The second term imposes regularization on $\phi(.)$. 

To guarantee the consistency of the image translation, the $\mathcal{L}_1$ loss is used to measure the difference between the warping of fake images on consecutive slice  $t^{th}$ and the translation of reference slice $k^{th}$. We define the temporal consistency loss function for mapping $X \to Y$ and $Y \to X$ as:
\begin{equation}
\begin{aligned}
&\mathcal{L}_{reg}(X, G_{X\to Y})&= \sum_{k=0, k \neq t}^{n} \left\lVert G_{X\to Y} (x_t) - G_{X\to Y} (x_k) \ocircle \phi(x_t, x_k)\right\lVert_{1} \\
&\mathcal{L}_{reg}(Y, G_{Y\to X})&= \sum_{k=0, k \neq t}^{n} \left\lVert G_{Y\to X} (y_t) - G_{Y\to X} (y_k) \ocircle \phi(y_t, y_k)\right\lVert_{1}
\end{aligned}
\label{eq:loss_reg}
\end{equation}
Fig. \ref{fig_deformation} illustrates an example for image-to-image translation from domain $X \to Y$ using temporal constraints. Let $x_t, x_{t+1}, x_{t+2}$ be consecutive slices of real images in the source domain $X$. A mapping function $G_{X \to Y}$ generates the fake image $y_t, y_{t+1}, y_{t+2}$ on target domain $Y$. On the source domain, we can learn displacement fields $\phi_t(.), \phi_{t+2}(.)$ between  $(x_t, x_{t+1})$ and $(x_{t+2}, x_{t+1})$. To constrain the consistency of the mapping from $X \to Y$, we minimize the distance (i) between the warped fake image $y_t \ocircle \phi_t(.)$ and $y_{t+1}$ for mapping from $t^{th}$ slice and $(t+1)^{th}$ slice, and (ii) between the warped fake image $y_{t+2} \ocircle \phi_{t+2}(.)$ and $y_{t+1}$ for mapping from $(t+2)^{th}$ slice and $(t+1)^{th}$ slice.    

\subsection{Network diagram}
\begin{figure}[t]
\centering
\includegraphics[width=1.0\textwidth]{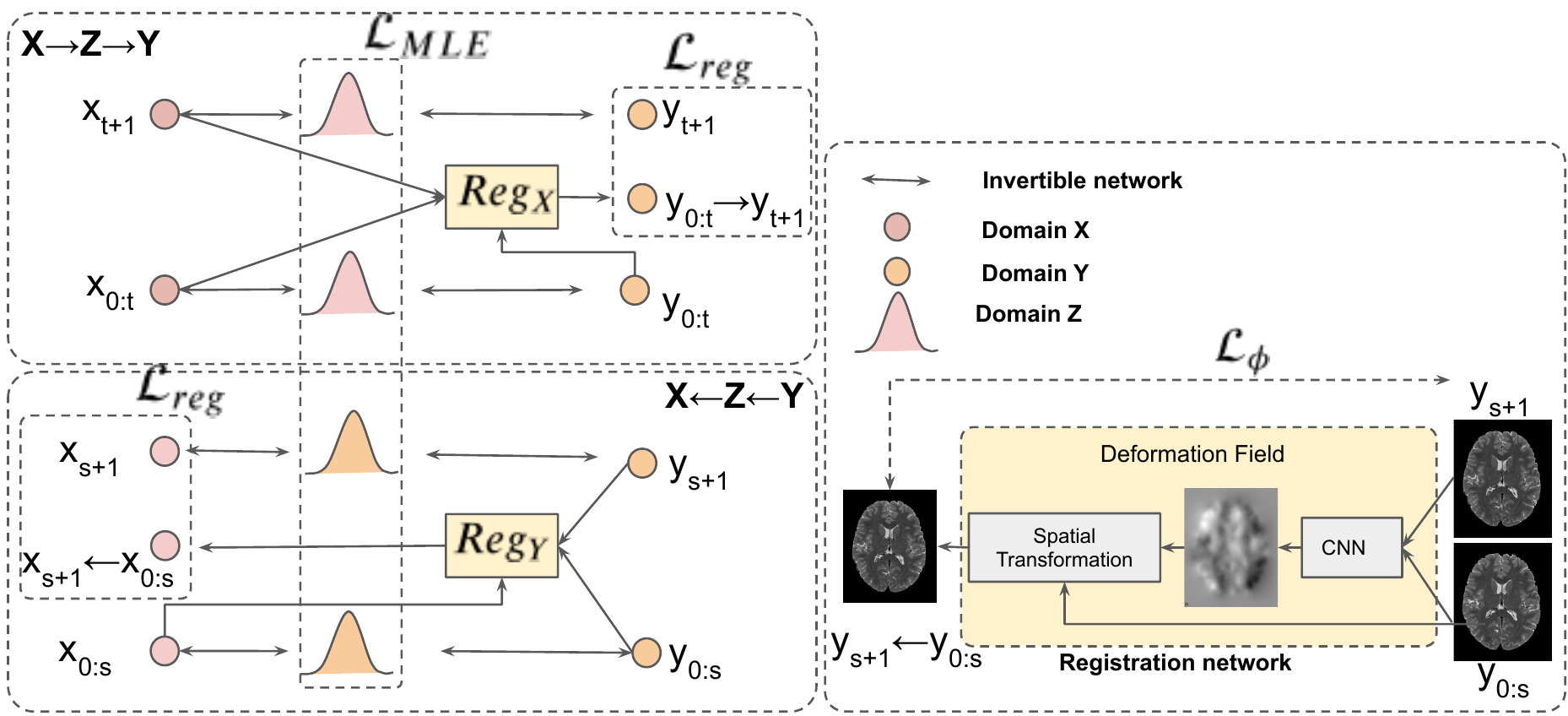}
\caption{Our flow-based deformation guidance approach for unpaired image-to-image translation.}
\label{fig_network_diagram}
\vspace{-0.5em}
\end{figure}
Fig. \ref{fig_network_diagram} illustrates the proposed network diagram for unpaired image-to-image translation. Our proposed network architecture inherits the advantages of invertible property of alignflow \cite{grover2019alignflow}. During training, we add two additional networks $Reg_X$ and $Reg_Y$ for each domain to learn the deformation field $\phi(.)$. These additional networks only use in training time, without increasing the model complexity and inference time comparison with the baseline flow-based method. The temporal constraint via $\mathcal{L}_{reg}(.)$ losses ensures the mapping of consecutive slices on the source domain should be consistent on the target domain. Finally, our objective function is defined as:
\begin{equation}
\begin{split}
&\mathcal{L}_{flow\_reg}(G_{X \to Y},G_{Y \to X},D_X,D_Y, \phi)= \mathcal{L}_{flow}(G_{X \to Y},G_{Y \to X},D_X,D_Y)
\\
&\quad\quad\quad\quad+\lambda_{1}\mathcal{L}_{reg}(X, G_{X\to Y})+\lambda_{2}\mathcal{L}_{reg}(Y, G_{Y\to X}) + \beta_{1}\mathcal{L}_X(\phi) + \beta_{2}\mathcal{L}_Y(\phi)\\
&\quad\quad\quad\quad+\gamma_1 \mathcal{L}_{TV}(X)
+ \gamma_2 \mathcal{L}_{TV}(Y)
\end{split}
\label{eq:loss_flow_based}
\raisetag{13pt}
\end{equation}
where  $\lambda_{1}$,  $\lambda_{2}$, $\beta_{1}$, and $\beta_{2}$ control the relative importance of the temporal consistence losses and the two registration losses. $\mathcal{L}_{TV}$ denotes total variation (TV) loss to impose spatial smoothness by measuring the horizontal
and vertical gradient of generated images \cite{yuan2018unsupervised}. These $TV$ losses are weighted by $\gamma_1, \gamma_2$.

\section{Experimental results}
\subsection{Datasets and Training}
We used common medical datasets to measure the robustness of our method against the existing methods: cycleGAN \cite{zhu2017unpaired}, recycleGAN \cite{bansal2018recycle}, cycleflow \cite{shen2020one} and alignflow \cite{grover2019alignflow}. cycleGAN \cite{zhu2017unpaired} is an unpaired image-to-image translation that works on single slice level. RecycleGAN \cite{bansal2018recycle} built upon the cycleGAN and add a temporal predictor that is trained to predict future slice in a set of previous consecutive slices. cycleflow \cite{shen2020one} is a flow-based method, but ignores the shared latent space $Z$ (directly map from $X \to Y$, instead of $X \to Z \to Y$ as the alignflow method). The synthetic image from each method was quantitatively compared with the real paired image using the following performance metrics: mean squared error (MSE), peak signal-to-noise ratio (PSNR), and structural similarity index (SSIM).

\textbf{Human Connectome Project (HCP)} is provided by the Human Connectome project \cite{van2013wu}. We used T1 as the source domain and T2 as the target domain. We extract the axial view of T1/T2 images into 2D images. We split the 2D images into 1150 images for training set and 500 images for testing set.

\textbf{MRBrainS13:} \cite{mendrik2015mrbrains} contains 15 subjects for training and validation and 6 subjects for testing. For each subject, two modalities are available that include T1-weighted, and T2-FLAIR with an image size of $48 \times 240 \times 240$.  We extract the dataset into 2D images with 450 images for training and 150 images for testing 

\textbf{Brats2019:} \cite{Brats} includes 210 HGG scans and 75 LGG scans. Each scan has a dimension of $240 \times 240 \times 155$. For each scan, we extract it to 2D images and use 770 images for training and 250 images for testing.

\textbf{Training}
All networks were implemented using the Pytorch framework and trained  on the 12GB GPU. The input image is resized to  $128 \times 128$ and normalized to $[-1, 1]$. We used axial slices (10 slices around the middle slice) from the each subject.  The Adam optimizer with a batch size of two was used to train the network. The initialization learning rate was set as 0.0002 and was decreased ten times every 20 epochs. We trained each model for 100 epochs. The balance weights were set as $\lambda_X=\lambda_Y=1e^{-5}, \lambda= \lambda_{1}=\lambda_{2}=10, \beta_1=\beta_2=1,\gamma_1=\gamma_2=1$. The discriminator network is a $70 \times 70$ PatchGAN \cite{isola2017image}. For alignflow network \cite{grover2019alignflow}, we set the number of scale was $1$, number of block was $3$. We use two consecutive slices (before and later slices) to learn the temporal constraint.

\subsection{Performance Evaluation}
\subsubsection{Qualitative evaluation}
\begin{figure}
\centering
\includegraphics[width=0.96\textwidth]{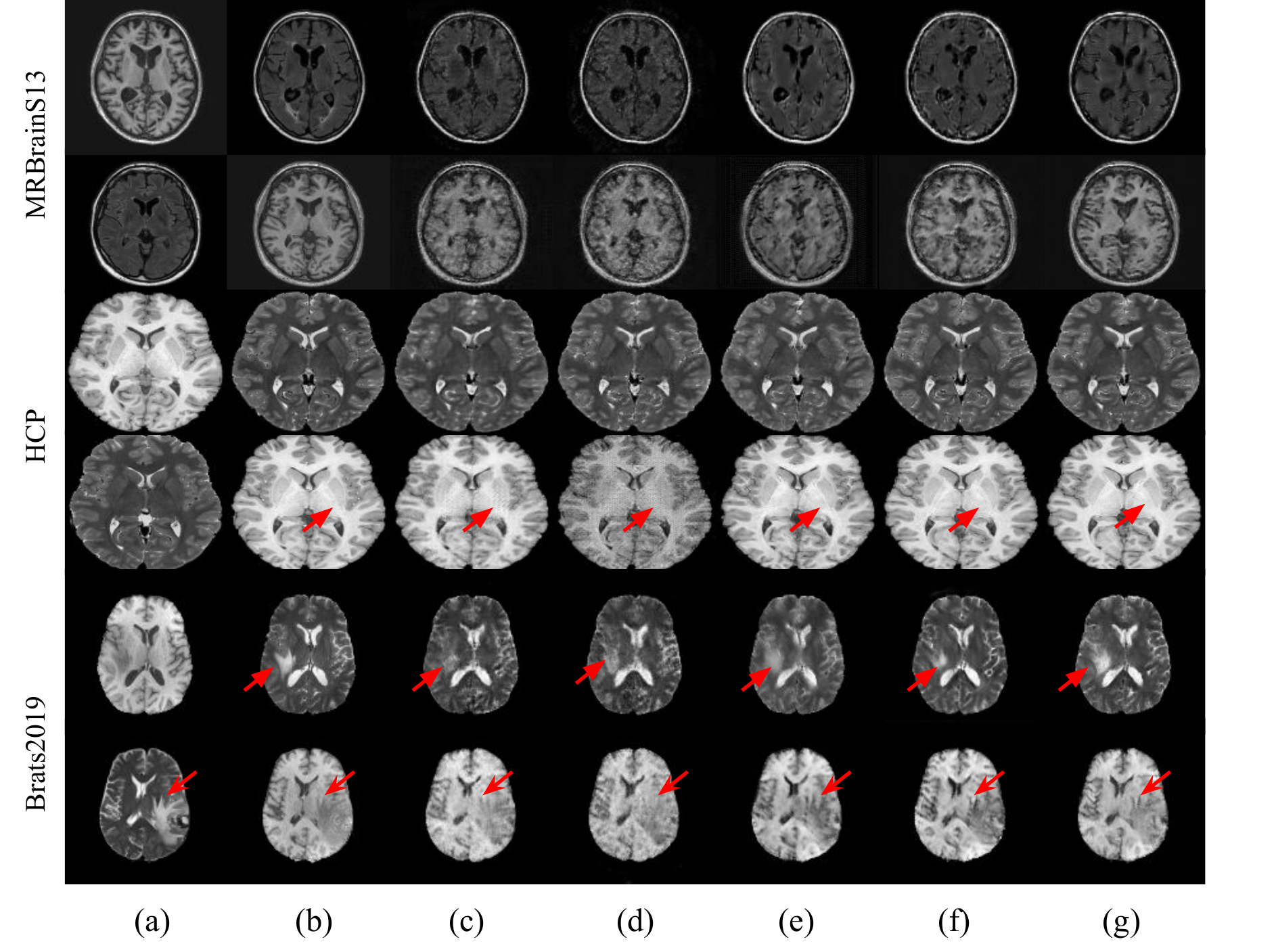}
\caption{A visualization of synthetic images on different datasets generated by (a) source image, (b) target image, (c) cycleGAN, (d) recycleGAN, (e) cycleflow, (f) alignflow, and (g) our method. Our method provides a good boundary on the tumor regions (red arrows in the fifth row) compared with the existing methods}
\label{fig_syn}
\end{figure}
Fig. \ref{fig_syn} illustrates the image translation on different datasets. The proposed methods (in the last column) provided a better synthetic image, resulting in better MSE, SSIM and PSNR scores. For example, the proposed synthetic T2 image provides a high qualitatively difference along the tumor boundary (indicated by the red arrows in the fifth row) than in existing methods using the available source T1 image as input. 

\subsubsection{Quantitative evaluation}
\begin{table}
\caption{Comparison between the proposed method against other image-to-image translation methods on \textbf{HCP, MRBrainS13, Brats19} datasets.}
\begin{tabular}{ll|c|c|c|c|c|c}
\toprule
\multicolumn{1}{c}{\bfseries} & \multicolumn{1}{c|}{\bfseries Method} & \multicolumn{2}{c|}{\bfseries MSE $\downarrow$} & \multicolumn{2}{c|}{\bfseries PSNR $\uparrow$} & \multicolumn{2}{c}{\bfseries SSIM $\uparrow$} \\
\cline{3-8}
& & T1 $\to$ T2 & T2 $\to$ T1 & T1 $\to$ T2 & T2 $\to$ T1 & T1 $\to$ T2 & T2 $\to$ T1 \\
\cline{1-8}
\parbox[t]{2mm}{\multirow{5}{*}{\rotatebox[origin=c]{90}{\underline{HCP}}}}
&\quad cycleGAN \cite{zhu2017unpaired} & 0.0193& 0.0167 & 23.2 & 24.4 & 0.783 & 0.793\\
&\quad recycleGAN \cite{bansal2018recycle} & 0.0212 & 0.0182 & 22.8 & 24.0 & 0.773 & 0.797\\
&\quad cycleflow \cite{shen2020one}  & 0.0213 & 0.0189 & 22.8 & 23.8 & 0.771 & 0.785\\
&\quad alignflow (baseline) \cite{grover2019alignflow} &  0.0200 & 0.0158 & 23.1 & 24.6 & 0.785 & 0.811\\
&\quad our method & \textbf{0.0179} & \textbf{0.0143} & \textbf{23.5} & \textbf{25.1} & \textbf{0.80} & \textbf{0.820}\\
\midrule
\parbox[t]{2mm}{\multirow{5}{*}{\rotatebox[origin=c]{90}{\underline{MRBrainS13}}}} 
& \quad cycleGAN \cite{zhu2017unpaired} & 0.0139 & \textbf{0.0235} & 24.7 & 22.4 & 0.793 & 0.704\\
& \quad recycleGAN \cite{bansal2018recycle} & 0.0154 & 0.0250 & 24.3 & 22.1 & 0.761 & 0.714\\
& \quad cycleflow \cite{shen2020one}  & 0.0158 & 0.0406 & 24.2 & 20.0 & 0.790 & 0.506\\
& \quad alignflow (baseline) \cite{grover2019alignflow} &  0.0165 & 0.0254 & 24.0 & 22.0 & 0.781 & 0.728\\
& \quad our method & \textbf{0.0128} & 0.0236 & \textbf{25.1} & \textbf{22.4} & \textbf{0.819} & \textbf{0.741}\\
\midrule
\parbox[t]{2mm}{\multirow{5}{*}{\rotatebox[origin=c]{90}{\underline{Brats2019}}}}
&\quad cycleGAN \cite{zhu2017unpaired} & \textbf{0.0178} & 0.0281 & \textbf{24.1}&22.7 & 0.833 &0.797\\
&\quad recycleGAN \cite{bansal2018recycle}& 0.0190 & 0.0272 & 23.8&22.6 &0.824 & 0.785\\
&\quad cyclelow \cite{shen2020one}  & 0.0251 & 0.0304 & 22.7&21.8 & 0.800 & 0.788\\
&\quad alignflow (baseline) \cite{grover2019alignflow} &0.022 & 0.0306 & 23.4&21.8 & 0.830 & 0.784\\
&\quad our method & 0.0188& \textbf{0.0258} & 23.9& \textbf{22.8} & \textbf{0.842} & \textbf{0.808}\\
\bottomrule
\end{tabular}
\label{table_mse}
\end{table}
Tables ~\ref{table_mse} reports the MSE, PSNR and SSIM values of the proposed method and existing methods. From the table, it is clear that the flow-based method (such as cycleflow \cite{shen2020one}, alignflow \cite{grover2019alignflow} and our method) provides competitive results with GAN-based method (such as cycleGAN, recycleGAN). By adding temporal constraints, the proposed network outperforms the baseline method (alignflow) on all performance metrics. Different from recycleGAN, that exploits temporal information via future slice prediction from consecutive slices, the proposed method measures pixel-wise temporal consistency by directly warping the synthetic slices with the deformation field of the consecutive slices from the source,  and thus achieves better performances. This indicates the effectiveness of the proposed method in the unpaired image to image translation for medical image.

\section{Conclusion}
We presented an effective method for image-to-image translation based on flow-based methods and deformation information that allows the proposed method to exploit the temporal information between consecutive slices to constrain the translation image. We show that the proposed method can provide a good translation image, yielding a better MSE, PSNR, and SSIM on various MRI datasets. Although our network is a fully invertible property, it requires more memory resource than GAN-based methods (such as cycleGAN, recycleGAN,...).


\end{document}